# Magnetic anisotropy determination and magnetic hyperthermia properties of small Fe nanoparticles in the superparamagnetic regime.

B. Mehdaoui, A. Meffre, L.-M. Lacroix, J. Carrey[*], S. Lachaize, M. Respaud

Université de Toulouse; INSA; UPS; LPCNO (Laboratoire de Physique et Chimie des Nano-Objets), 135 avenue de Rangueil, F-31077 Toulouse, France and
CNRS; UMR 5215 ; LPCNO, F-31077 Toulouse, France

M. Gougeon

Institut CARNOT - CIRIMAT - UMR 5085, Bâtiment 2R1, 118 route de Narbonne
F-31062 Toulouse, France

B. Chaudret

Laboratoire de Chimie de Coordination-CNRS, 205 rte de Narbonne, 31077 Toulouse cedex 4, France

**Abstract:**

We report on the magnetic and hyperthermia properties of iron nanoparticles synthesized by organometallic chemistry. They are 5.5 nm in diameter and display a saturation magnetization close to the bulk one. Magnetic properties are dominated by the contribution of aggregates of nanoparticles with respect to individual isolated nanoparticles. Alternative susceptibility measurements are been performed on a low interacting system obtained after eliminating the aggregates by centrifugation. A quantitative analysis using the Gittleman's model allow a determination of the effective anisotropy $K_{eff} = 1.3 \times 10^5$ J.m$^{-3}$, more than two times the magnetocristalline value of bulk iron. Hyperthermia measurements are performed on agglomerates of nanoparticles at a magnetic field up to 66 mT and at frequencies in the range 5-300 kHz. Maximum measured SAR is 280 W/g at 300 kHz and 66 mT. Specific absorption rate (SAR) displays a square dependence with the magnetic field below 30 mT but deviates from this power law at higher value. SAR is linear with the applied frequency for $\mu_0 H$=19 mT. The deviations from the linear response theory are discussed. A refined estimation of the optimal size of iron nanoparticles for hyperthermia applications is provided using the determined effective anisotropy value.

**Main Text:**

### I. INTRODUCTION

Magnetic nanoparticles (MNPs) are of particular interest for biomedical application such as single molecule detection [1], drug release [2] or magnetic hyperthermia treatment [3]. In order to lower the detection limit or to provide an important temperature increase, all these applications require MNPs with a high magnetic moment [4]. Therefore, the widely used iron oxides particles, such as maghemite (γ-Fe$_2$O$_3$, 80 Am$^2$kg$^{-1}$) or magnetite (Fe$_3$O$_4$, 120 Am$^2$kg$^{-1}$), could be advantageously replaced by metallic particles such as cobalt (Co, 160 Am$^2$kg$^{-1}$), iron (Fe, 220 Am$^2$kg$^{-1}$), or iron cobalt alloy (FeCo, 240 Am$^2$kg$^{-1}$).

Moreover, their size must be optimised to maximize the specific absorption rate (SAR) of NPs. The optimal size strongly depends on their magnetic anisotropy since both size and anisotropy directly influences their coercive field [5,6]. For instance, it has been estimated that the optimal size of iron nanoparticles for magnetic hyperthermia experiments performed at 100 kHz and 20 mT is around 15 nm, considering magnetically independent NPs with bulk anisotropy [6]. However, this value may be adapted depending on the real anisotropy value. Accurate measurement of the magnetic anisotropy of MNPs is not trivial. One possibility is the precise alternative susceptibility characterizations on well diluted assemblies of magnetically independent nanoparticles.

Our group has developed an organometallic synthesis allowing the controlled growth of pure metallic iron NPs displaying the bulk magnetization [7,8,9]. So far, only the determination of the magnetic anisotropy of 1.5 nm in diameter MNPs has been determined and was found equal to $5.2 \times 10^5$ J /m$^3$, *i.e.* ten times the bulk value [10]. In this article, we perform a first step toward bigger NPs by the careful study of 5.5 nm NPs displaying a magnetization close to the bulk one. Hyperthermia measurements as a function of the magnetic field and of the applied frequency have been performed on samples in which the NPs are in strong magnetic interactions. These results are discussed and some conclusions on the optimal size of iron nanoparticles for magnetic hyperthermia are provided.

## II. RESULTS

Iron nanoparticles exhibiting a polycrystalline bcc structure were synthesized by an organometallic approach in mesitylene solvent [8]. The resulting "raw" solution is composed of both isolated 5.5 nm NPs and of micrometric agglomerates, as evidenced by both TEM micrographs [see Fig.1(a)] and in dynamic light scattering (DLS) measurements [see Fig.1(c)]. The size of the NPs inside the agglomerates is roughly the same as the one of the isolated NPs. These aggregates can be removed from the colloidal solution using a standard centrifugation process (20 min at 25.000 rpm). In order to prevent any oxidation, the samples were always kept under an Ar atmosphere. After centrifugation, the supernatant contains only dispersed nanoparticles as evidenced by the DLS results [see Fig.1(d)]. A single hydrodynamic size distribution is then observed, centred on 8 nm, which is in good agreement with the mean diameter of the particles and their surrounding $C_{16}$ surfactants. This result is confirmed by susceptibility measurements on the "raw" solution and on the supernatant, following a zero field cooling (ZFC) / field cooling (FC) procedure [see Figs.1(e) and 1(f)]. The ZFC curve on the "raw" solution reveals two broad maxima which can be seen as the signature of i) dispersed NPs with a blocking temperature ($T_B$) of 30 K, and ii) aggregates of MNPs in which dipolar interactions took place ($T_B$ = 136 K). On the other hand, a unique maxima at 31 K remains in the ZFC of the supernatant after centrifugation. In this case, the magnetic susceptibility follows a Curie-Weiss law $\chi \propto (T-\theta)^{-1}$ for $T>T_B$, with $\theta$ = -5.0 K (not shown).

The saturation magnetization ($M_S$) measured at 2 K for both type of NPs ($M_S$ = 207 A.m$^2$.kg$^{-1}$) is close to the bulk iron magnetization, i.e., 212 A.m$^2$.kg$^{-1}$ (data not shown here). Moreover the remnant magnetization ($M_R$) increases from 41 to 50 % of $M_S$ after the centrifugation process, which is the expected value predicted by the Stoner-Wohlfarth theory for non-interacting particles [11].

Frequency dependent susceptibility measurements $\chi_{AC}$ were performed on this colloidal solution. The temperature dependence of the in-phase ($\chi'$) and the out-of-phase ($\chi''$) components of the susceptibility were measured for frequencies ranging from 0.1 to 1500 Hz. Figure 2 displays the experimental data and the fitting using Gittleman's model: [12]

$$\overline{\chi}(T,\omega) = \frac{\int_0^\infty \widetilde{\chi}_V(T,\omega) v f(v) dv}{\int_0^\infty v f(v) dv} \qquad (1)$$

with $\widetilde{\chi}_V(T,\omega) = \frac{\chi_S(T) + i\omega\tau\chi_b(T)}{1 + i\omega\tau}$. $\chi_S = \frac{\mu_0 M_S^2(T) v}{3 k_B T}$ and $\chi_b = \frac{\mu_0 M_S^2(T)}{3 K_{eff}}$ are respectively the contributions from superparamagnetic and blocked nanoparticles. While the saturation magnetization has been fixed according to the magnetization measurements, the other parameters, like the log normal size distribution $f(v)$ characterized by the mean volume $v$ and its width $\sigma$, the relaxing time $\tau_0$ and the effective anisotropy constant $K_{eff}$ are extracted from the $\chi_{AC}$ curves. Their values are summarized in Table 1.

Magnetic hyperthermia has been measured as a function of the applied magnetic field amplitude and frequency on a concentrated solution of the "raw" nanoparticles. For frequencies up to 100 kHz, hyperthermia experiments were performed on a home-made frequency-adjustable electromagnet with a magnetic field ranging up to 30 mT [13]. For $f$ = 300 kHz, the measurements were performed on an induction oven working at a fixed frequency and a maximum magnetic field of 66 mT. An ampoule containing the colloidal solution was sealed under vacuum to prevent any oxidation of the NPs. The ampoule was placed in a calorimeter with 1.5 ml of water, the temperature of which is measured. More details on the measurement method are reported in Refs [5, 6].

SAR as a function of magnetic field in the range 0-30 mT, measured at frequencies of 50, 100 and 300 kHz, are shown in Fig. 3(a). At low field, the SAR shows a square dependence versus the magnetic field, in agreement with the linear response theory for superparamagnetic particles. Fig. 3(b) displays the SAR values measured at 300 kHz as function of the applied magnetic in the range 0-66mT. For a field above 30 mT, the SAR increases sharply with increasing field, and is fitted by a power law function $SAR \sim H^{3.2}$. The SAR measured at 66 mT is 280 W/g, much weaker than the one measured in the same conditions on 16 nm nanocubes [6]. Fig. 3(c) displays the frequency dependence of the SAR at $\mu_0 H$ = 19.3 mT. A clear linear dependence of SAR as a function of the frequency was observed.

### III. DISCUSSION

Although the aggregates, composed of nanoparticles under strong dipolar interactions, are efficiently precipitated out by the centrifugation process, the NPs remaining in the colloidal solution still interacts with each others. These interactions are relatively weak ($\theta$ = 5 K) and do not affect the hysteresis cycle feature ($M_R/M_S$ = 0.5). However, they drastically influence the dynamical properties leading to a very short intrawell relaxation time $\tau_0 = 10^{-15}$s. Such a low value, compared to the usually reported ones ($10^{-9}$ - $10^{-10}$s) for individual nanoparticle is consistent with previous experiments on interacting NPs reported by Dorman *et al.* [14]. The fitting of the $\chi_{AC}$ using the $M_S$ value (207 Am²kg⁻¹) deduced from magnetization measurement gives an estimate of the mean diameter (5.1 nm) in agreement with the TEM pictures. The effective anisotropy constant ($K_{eff}$ = 1.3 × 10$^5$ J.m$^{-3}$) is more than two times higher than the magnetocristalline value of bulk bcc iron ($K_{MC}$ = 4.8 × 10$^4$ J.m$^{-3}$). Due to the weak value of $\theta$, this increase is rather attributed to size reduction effects [15] than to the presence of magnetic interactions.

Hyperthermia measurements are performed on samples where a large majority of the NPs are in aggregates, displaying hydrodynamic radii $r$ larger than 250 nm. The Brownian frequency $f_B$ of these agglomerates, calculated using $f_B = k_B T / 4\pi\eta r^3$, where $\eta$ = 0.69 10$^{-3}$ kg m$^{-1}$ s$^{-1}$ is the viscosity coefficient of the solvent, leads to $f_B$ = 240 Hz. Therefore, the agglomerates, and thus the NPs inside them, can be regarded as fixed in the range 2-300 kHz. Consequently, the losses observed cannot be explained by the mechanical rotation of the NPs themselves (Brownian relaxation), but rather by the reversal of the magnetization inside the nanoparticles.

Fig. 3 shows that below a field $\mu_0 H$ = 30 mT, the variation of SAR as a function of $\mu_0 H$ follows a square law, in agreement with the linear response theory (LRT) for superparamagnetic particles [16]. However, for $\mu_0 H$ > 30 mT, this power law increases to an exponent 3.2, which is not predicted by the LRT. Moreover, the linear variation of the SAR as a function of the frequency at $\mu_0 H$ = 19.3 mT is also in disagreement with the LRT, which predicts a non-linear dependency with frequency. Attempts to fit the frequency dependence of the SAR using the LRT and taking into account a size distribution and an effective anisotropy distribution of the nanoparticles failed.

We think of two reasons for which the LRT might fail: i) above a given magnetic field the LRT is not valid anymore, and/or ii) the magnetic interactions influence too much the response of the system, even at small magnetic fields. With respect to the first hypothesis, the LRT is expected to be valid for magnetically independent NPs only when $\frac{\mu_0 H M_s V}{k_B T} < 1$ (this is the condition for which a Langevin function is linear with the magnetic field) and when $\mu_0 H << \frac{2K}{Ms}$ (this is the condition for which the Néel relaxation time is independent of the magnetic field). The two conditions leads to $\mu_0 H$ < 300 mT and $\mu_0 H$ << 160 mT in our system, respectiviely. While the first condition is clearly fulfilled, we cannot exclude that the deviation from the LRT above 30 mT could be due to the non-respect of the second one. However, these considerations cannot explain the linearity of the frequency-dependence of the SAR at 19.3 mT; we think that this behaviour is a consequence of the presence of magnetic interactions. Clearly, theoretical results on the hyperthermia properties of interacting superparamagnetic MNPs are highly desirable to be compared with such experimental results.

In the context of hyperthermia applications, the determination of the effective anisotropy of iron MNPs permit to refine the estimation of the diameter maximising the SAR value. In a previous article, we showed that the largest SAR should be obtained for MNPs in the ferromagnetic regime; we estimated that the diameter maximizing the SAR should be around 15 nm for magnetically independent MNPs with bulk anisotropy [6]. However, when performing the same calculation but using the anisotropy $K_{eff}$ = 1.3 × 10$^5$ J.m$^{-3}$ determined here, a diameter around 9.5 nm for $\tau_0$=10$^{-10}$s is obtained. NPs bigger than 5.5 nm should probably display an anisotropy intermediate between the one determined here and the one of the bulk, so the diameter maximizing the SAR of magnetically independent iron nanoparticles should be found somewhere between 9.5 and 15 nm, depending on their $K_{eff}$ value. Hyperthermia experiments on magnetically independent MNPs are desirable to check the validity of such predictions.

## Acknowledgements:

We acknowledge InNaBioSanté foundation, AO3 program from Université Paul Sabatier (Toulouse) and Conseil Régional de Midi-Pyrénées for financial support, V. Collière (TEMSCAN) for TEM, C. Crouzet for technical assistance and A. Mari for magnetic measurements.

**Table:**

| $d_{moy}$ (nm) | Distribution | $M_S$ (Am$^2$kg$^{-1}$) | $\tau_0$ (s) | $K_{eff}$ (J.m$^{-3}$) |
|---|---|---|---|---|
| 5.1 | Log-Normal $\sigma = 0.138$ | 207 | $10^{-15}$ | $1.3 \times 10^5$ |

Table 1. Adjustment parameters extracted from the fitting of the experimental $\chi_{AC}$ curves with Gittleman's model.

**Figures:**

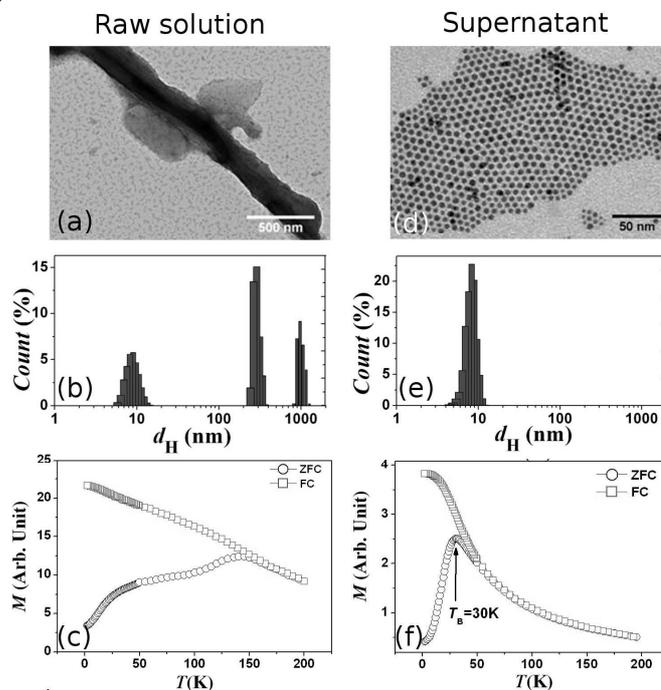

Figure 1. (a-b) TEM micrographs (c-d) hydrodynamic size distribution extracted from DLS measurements and (e-f) DC magnetic susceptibility (ZFC-FC) measurements. (a,c,e) correspond to "raw" solutions and (b,d,f) to centrifugated NPs.

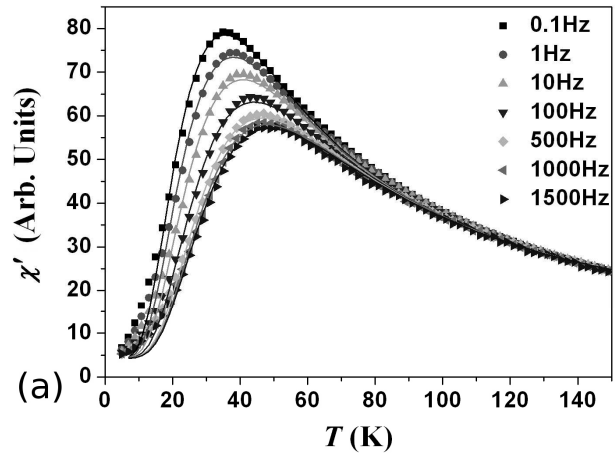

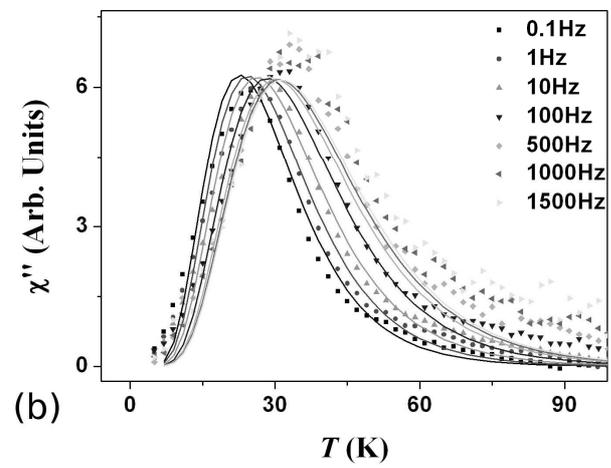

Figure 2. Temperature dependence of the (a) in-phase ($\chi'$) and (b) out-of-phase ($\chi''$) components of the alternative susceptibility for varying frequencies. Symbols represent experimental data, lines are the fits from Gittleman's model using the parameters shown in Table 1.

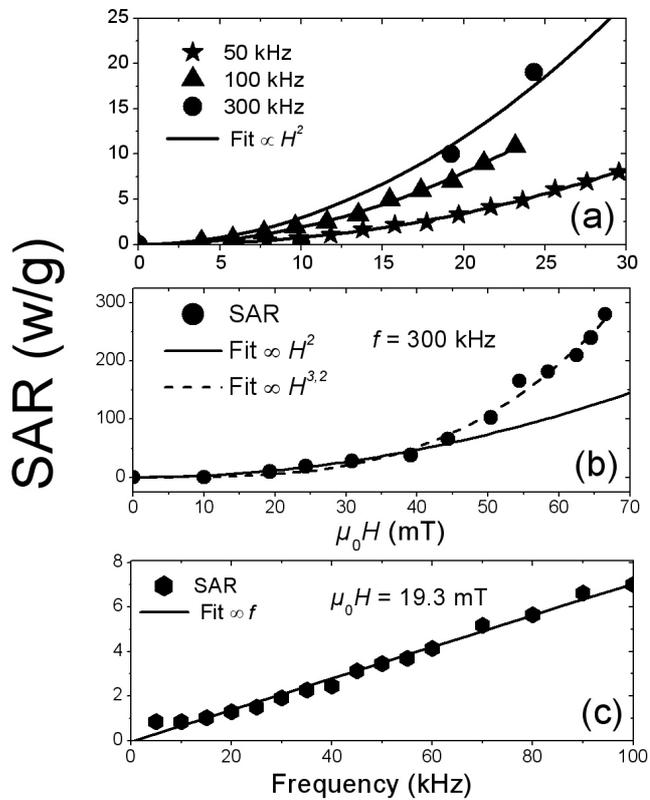

Figure 3: (a) SAR as a function of magnetic field in the range [0-30 mT], measured at $f$ = 50, 100 and 300 kHz, fitted using SAR ~ $H^2$. (b) SAR as a function of magnetic field in the range [0-66 mT], measured at $f$ = 300 kHz, fitted using SAR ~ $H^2$ and $H^{3.2}$. (c) Evolution the SAR as a function of frequency, measured at $\mu_0 H$ = 19.3 mT, fitted using a linear function.